\documentclass[10pt, twocolumn, notitlepage, superscriptaddress, prb, longbibliography]{revtex4-2}

\usepackage{here}
\usepackage{amsmath}         
\usepackage{graphicx}
\usepackage{braket}         
\usepackage{lettrine}
\usepackage{color}
\usepackage{hhline}
\usepackage{tabularx}
\usepackage{calc}
\usepackage{soul}

\begin{document}

\title{Spin and orbital transport in rare earth dichalcogenides: The case of EuS$_2$} 

\author{Mahmoud Zeer}
\email{m.zeer@fz-juelich.de}
\affiliation{Peter Gr\"unberg Institute and Institute for Advanced Simulation, Forschungszentrum J\"ulich and JARA, 52425 J\"ulich, Germany}
\affiliation{Department of Physics, RWTH Aachen University, 52056 Aachen, Germany}

\author{Dongwook Go}
\email{d.go@fz-juelich.de}
\affiliation{Peter Gr\"unberg Institute and Institute for Advanced Simulation, Forschungszentrum J\"ulich and JARA, 52425 J\"ulich, Germany}
\affiliation{Institute of Physics, Johannes Gutenberg-University Mainz, 55099 Mainz, Germany}

\author{Johanna P. Carbone}
\affiliation{Peter Gr\"unberg Institute and Institute for Advanced Simulation, Forschungszentrum J\"ulich and JARA, 52425 J\"ulich, Germany}
\affiliation{Department of Physics, RWTH Aachen University, 52056 Aachen, Germany}

\author{Tom G. Saunderson}
\affiliation{Institute of Physics, Johannes Gutenberg-University Mainz, 55099 Mainz, Germany}

\author{Matthias Redies}
\affiliation{Peter Gr\"unberg Institute and Institute for Advanced Simulation, Forschungszentrum J\"ulich and JARA, 52425 J\"ulich, Germany}
\affiliation{Department of Physics, RWTH Aachen University, 52056 Aachen, Germany}

\author{Mathias Kl\"aui}
%\email{y.mokrousov@fz-juelich.de}
%\affiliation{Peter Gr\"unberg Institute and Institute for Advanced Simulation, Forschungszentrum J\"ulich and JARA, 52425 J\"ulich, Germany}
\affiliation{Institute of Physics, Johannes Gutenberg-University Mainz, 55099 Mainz, Germany}

\author{Jamal Ghabboun}
\affiliation{Department of Physics, Bethlehem University, Bethlehem, Palestine}

\author{Wulf Wulfhekel}
\affiliation{Institute of Physics, Johannes Gutenberg-University Mainz, 55099 Mainz, Germany}

\author{Stefan Bl\"ugel}
\affiliation{Peter Gr\"unberg Institute and Institute for Advanced Simulation, Forschungszentrum J\"ulich and JARA, 52425 J\"ulich, Germany}

\author{Yuriy Mokrousov}
\email{y.mokrousov@fz-juelich.de}
\affiliation{Peter Gr\"unberg Institute and Institute for Advanced Simulation, Forschungszentrum J\"ulich and JARA, 52425 J\"ulich, Germany}
\affiliation{Institute of Physics, Johannes Gutenberg-University Mainz, 55099 Mainz, Germany}

\begin{abstract}
We perform first-principles calculations to determine the electronic, magnetic and transport properties of rare-earth dichalcogenides  taking a monolayer of the H-phase  EuS$_2$ as a representative. We predict that the H-phase of the EuS$_2$ monolayer exhibits a half-metallic behavior upon doping with a very high magnetic moment.
%of $6.9\ \mu_B$ per unit cell and a large band gap. 
We find that the electronic structure of EuS$_2$ is very sensitive to the value of Coulomb repulsion $U$, which effectively controls the degree of hybridization between Eu-$f$ and S-$p$ states. We further predict  that the non-trivial electronic structure of EuS$_2$ directly results in a pronounced anomalous Hall effect with non-trivial band topology. Moreover, while we find that the spin Hall effect closely follows the anomalous Hall effect in the system, the orbital complexity of the system results in a very large orbital Hall effect, whose properties depend very sensitively on the strength of correlations. Our findings thus promote rare-earth based dichalcogenides as a promising platform for topological spintronics and orbitronics. 
%e the transport calculations of H-EuS$_2$ by means of the band interpolation using Maximally localized Wannier function. Our finding predicts that the orbital Hall conductivity (OHC) to be much larger than spin Hall conductivity (SHC) as a function of Fermi energy position, but both vanish in the band gap. On another hand, Anomalous Hall conductivity has a large magnitude as well that promotes H-EuS$_2$ as a 
%promising platform for sizeable anomalous Hall effect.
\end{abstract}

\date{\today}                 
\maketitle

\section{Introduction}
Two-dimensional (2D) transition-metal dichalcogenides (TMDs) have attracted enormous attention due to their strong potential as a platform for various flavors of spintronics effects. The rich set of electronic phases that TMDs exhibit includes semiconductors~\cite{Splendiani2010,Gutierrez2013,Miro2014}, semimetals~\cite{Lee2015}, metals~\cite{Zhao2016}, and superconductors~\cite{Guillamn2011}, which allows for novel design paradigms incorporating multiple complex phases of matter.
Furthermore, due to a combination of low crystal symmetry and spin-orbit coupling (SOC)~\cite{Zhu2011}, these materials can host exotic excitations and exhibit different transport properties. For example, recent studies suggest that TMDs naturally lacking inversion symmetry can be exploited to drive a variety of Hall effects such as the anomalous Hall effect (AHE)~\cite{Fuh2016}, valley Hall effect~\cite{Mak2014,Zhou2021}, spin Hall effect (SHE)~\cite{Feng2012,Xiao2012}, or even orbital Hall effect (OHE)~\cite{Cysne2021,Canonico2020,Bhowal2020,Canonico2020*,Liu2019,Bernevig2005,Tanaka2008,Jo2018,kontani2009,Go2018,cysne2022}. 
 
Among the latter two phenomena, SHE and OHE attract particular attention owing to bright prospects for spintronic applications~\cite{Shan2013,Bhowal2019, Bhowal2019*,You2019}. While the SHE is a well-known and by now in-depth studied phenomenon~\cite{Sinova2015}, the OHE is much less explored~\cite{Bhowal2019,Jo2018,Bernevig2005,Go2018,kontani2009}.
%Beyond this, due to the high SOC and low symmetry these materials exhibit, another Hall effect is making an appearance, the orbital Hall effect (OHE) \cite{Bhowal2019}. 
The essence of the OHE is in the coupling between the orbital motion of electrons and an electric field, which drives a transverse flow of orbital angular momentum, as opposed to spin angular momentum in SHE~\cite{Bernevig2005,Bhowal2020,Bhowal*2020,Cysne2021,Tokatly2010,Bhowal2021}. 
%In both, the flow of transverse spin(orbital) current are generated by applying a longitudinal electric field \cite{Bernevig2005}.
Such orbital Hall currents can for example generate a measurable effect via the generation of orbital type of torques on the magnetization~\cite{Go2020,Soogil2021}, or give rise to strong non-local spin currents via the effect of spin-orbit coupling~\cite{Go2020*}. This also gives rise to a new channel for information transfer $-$ the orbital channel. Harnessing the physics of orbital currents is expected to push spintronics into a new direction, the direction of  orbitronics~\cite{Go2021}.

On the side of materials, OHE has been studied predominantly in  non-magnetic crystals, and very little is known about orbital currents in magnetic systems, especially in low dimensions. In this light,  magnetism in 2D materials can be extremely profitable for  spintronics and orbitronics~\cite{Gibertini2019, Zhai2019,Shao2021}. 
% While it has long been believed that long range magnetism in 2D materials is impossible due to thermal fluctuations based on the 2D isotropic Heisenberg model by Mermin-Wagner theorem prediction \cite{Mermin1966}. 
Recent discoveries of magnetic order in various classes of 2D magnets such as CrI$_3$~\cite{Huang2017}, Fe$_3$GeTe$_2$~\cite{Deng2018}, and bilayer Cr$_2$Ge$_2$Te$_6$ \cite{Gong2017,Fengfeng2021} have given the exploration of novel 2D ferromagnetic materials and their magnetic properties a significant momentum. On the side of TMDs, several representative compounds, such as metallic VX$_2$ \cite{Hui2014,Ma2012,Wei2020}, CrX$_2$ \cite{Wei2020,Wang2018,Lv2015} and semiconducting MnS$_2$ (X=S, Se, Te) \cite{Ataca2012,Kan2014,Wei2020}, have been shown to exhibit magnetic ordering in their ground state.
%, with known examples being metallic VX$_2$ \cite{Hui2014,Ma2012,Wei2020}, CrX$_2$ \cite{Wei2020,Wang2018,Lv2015} and semiconducting MnS$_2$ (X=S, Se, Te) \cite{Ataca2012,Kan2014,Wei2020}. 
A very promising direction to pursue here is the realization of 2D magnetic materials, and in particular dichalcogenides, based on 4$f$-elements. Given a strong tendency to magnetism and strong SOC combined with orbital complexity of 4$f$-electrons, $4f$-based 2D materials are expected to serve as a fruitful foundation for various  spin and especially orbital transport effects. 
%Another interesting direction to achieve magnetism in 2D materials is to utilize $4f$ elements, where the interplay of the local Coulomb interaction, crystal field, and spin-orbit coupling may lead to novel physical phenomena. 

 %Naturally, following these discoveries a significant interest is being paid to investigating their anomalous and spin Hall effects \cite{Shan2013}, as these have enormous applications in spintroinc devices\cite{Shan2013,Bhowal2019, Bhowal2019*,You2019}. 
% Beyond this, due to the high SOC and low symmetry these materials exhibit, another Hall effect is making an appearance, the orbital Hall effect (OHE) \cite{Bhowal2019}. The orbital Hall effect is in analogy to the SHE, where the coupling between the orbital moment and the crystal drive an orbital (as opposed to spin) current in the material. In both, the flow of transverse spin(orbital) current are generated by applying a longitudinal electric field \cite{Bernevig2005}. The orbital current generates a measurable effect via the generation of a spin current through SOC \cite{Go2020}, sometimes giving rise to spurious spin currents which have magnitudes larger than anticipated. This gives rise to a new channel within which information can be carried, the orbital channel. Harnessing this will push the field of spintronics into a new direction, the direction of spin orbitronics.
 
In this work, we perform first-principles calculations of structural, electronic, magnetic and transport properties of  H-phase monolayer of the $4f$ rare-earth  dichalcogenide  EuS$_2$. Taking EuS$_2$ as an example, we thus aim to uncover the potential of the 4$f$ rare-earth  dichalcogenides as possible sources of pronounced charge, spin, and orbital currents. Keeping in mind the importance of correlation effects in $f$-materials, we consider a range of Coulomb strengths as given by the parameter $U$, finding that  EuS$_2$ has a very large magnetic moment and a large band gap for a wide range of correlation strengths. Importantly, we find a strongly increased $p-d-f$ hybridization among Eu and S atoms upon increasing $U$, which has a drastic influence on the transport properties.
%
%, implying this system to be a ferromagnetic semiconductor regardless of the strength of the electron-electron correlations. 
Specifically, our calculations predict that in the strongly-correlated limit the electronic bands in EuS$_2$ have a non-trivial topology, which results in sizable nearly-quantized values of the anomalous Hall conductivity (AHC) exhibited by the $f$-states. Remarkably, we find that the SHE in the system largely follows the behavior of the AHE owing to the strong majority character of the occupied $f$-states, while the OHE exhibits a very non-trivial behavior as a function of band filling, reaching a magnitude more than six times larger than that of the SHE. The detailed analysis we have carried out of the electronic structure in conjunction with Hall effects in the EuS$_2$ makes it possible to grasp the basic principles that can be used to turn rare-earth dichalcogenides into an an efficient source of orbital currents. Given a strong tendency of this class of materials to magnetism we thus propose that  rare-earth  dichalcogenides can occupy a unique place in the exotic niche of 2D magnetic materials with high potential for spintronics and orbitronics. %theof the Hall effects in this material Then, we carry out calculations of the intrinsic AHE, SHE and OHE for monolayer H-EuS$_2$ induced by applying an in-plane electric field. Not only to do we predict the appearance of a large AHE as a result of the existence of the magnetic ordering, but we also observe an OHE significantly larger than the SHE response. We believe this work will pave the way for the experimental discovery of a new family of 4$f$  rare earth metal dichalcogenide systems which will be instrumental to realising the potential of spin orbitronics. The rest of the paper is organized in the following way.
Our manuscript is structured as follows: in Section II we provide computational details of our study. In Section III we present  the results of our calculations and analysis, including a discussion, and we end with a brief conclusion in section IV.

\section{Computational Details}
Our first-principles calculations were performed using the film version of full-potential linearized augmented plane wave method ~\cite{Wimmer1981}, as implemented in the J\"ulich density functional theory (DFT) code FLEUR~\cite{fleur}.
We used the Perdew-Burke-Ernzerhof approximation to the exchange-correlation potential~\cite{Perdew1996}. The structure of the monolayer was relaxed so that residual forces are well below 0.001\,eV/\AA. 
For self-consistent calculations we used a 16$\times$16 Monkhorst-Pack grid in the first Brillouin zone and a plane-wave cutoff of $4.1$\,$a_0 ^{-1}$, where $a_0$ is the Bohr radius. %Plane-wave cutoffs for potential and exchange-correlation potential were chosen both to be $12.3 a_0^{-1}$. 
We set the angular momentum expansion of 10 and 8 for Eu and S atoms, respectively.
The muffin-tin radii of Eu and S atoms were set to $2.8$\,$a_0$ and $1.94$\,$a_0$, respectively. Spin-orbit coupling (SOC) was included  self-consistently within the second variation scheme for all calculations.

To treat the effect of strongly correlated electrons in the 4$f$-shell of Eu we applied the GGA+$U$ method within the self-consistent DFT cycle {\cite{Shick1999}. The on-site Coulomb interaction strength $U$ applied to the Eu 4$f$ states was varied from $0$\,eV to $6.7$\,eV, and the intra-atomic exchange interaction strength $J$ was chosen to be $0.7$\,eV~\cite{Shick1999}. 
%For total energy calculations, self-consistency was performed with an energy convergence of $10^{-6}\ \mathrm{eV}$. 
%The atomic structure of H-EuS$_2$ monolayer has the P6m2$_{D3h}$ ({\bf \color{blue} never saw this notation}) point group symmetry.
The optimized structure of H-EuS$_2$ is shown in Fig.~\ref{fig:atomic_structure} together with the definition of the axes. We first performed structural relaxations without $U$, which yielded a lattice constant of H-EuS$_2$ of 4.616\,\AA~and a distance between the planes of Eu and S atoms  along the $z$-axis of 1.1\,\AA. Adding a Coulomb repulsion strength of $U=6.7$\,eV modified the values of respective distances to 4.744\,\AA~and 1.08\,\AA. All calculations were performed for the latter values of structural constants. 

Transport calculations were performed for two values of $U$: 2.5\,eV and 6.7\,eV.  %\cite{Shick1999}, and $U_\text{eff}=2.5 \ \mathrm{eV}$ as well. 
The calculations of AHC, spin Hall conductivity (SHC), and orbital Hall conductivity (OHC) were performed according to expressions discussed below by exploiting the technique of Wannier interpolation~\cite{Pizzi2020,Freimuth2008}. In order to do this, 36 maximally-localized Wannier functions were constructed out of the {\it ab-initio} electronic structure, by projecting the Bloch states onto $p$-states of S and $f,d$-states of Eu atoms, so as to reproduce the electronic structure in an energy window that contains all the states below the Fermi energy and enough number of states above the Fermi energy. Throughout this work we treat the system as an perpendicularly magnetized ferromagnet. For $U=6.7$\, eV the magnetic anisotropy energy was estimated to be of the order of 30\,$\mu$eV favoring the out-of-plane
direction of magnetization, while the ferromagnetic state was found to be 0.3\,meV lower in energy than the row-wise antiferromagnetic state.  
%{\bf{\color{red} 36
% WFs is considered that contributed into the $P$ states of S and both $f$, $d$ states of Eu , while the frozen energy window is chosen to $+2 eV$ above the Fermi energy.}}. 

%{\bf {\color{blue} Need a bit more details on the number of WFs, projections, frozen window.}}

\begin{figure}[t!]
    \includegraphics[angle=0, width=0.4\textwidth]{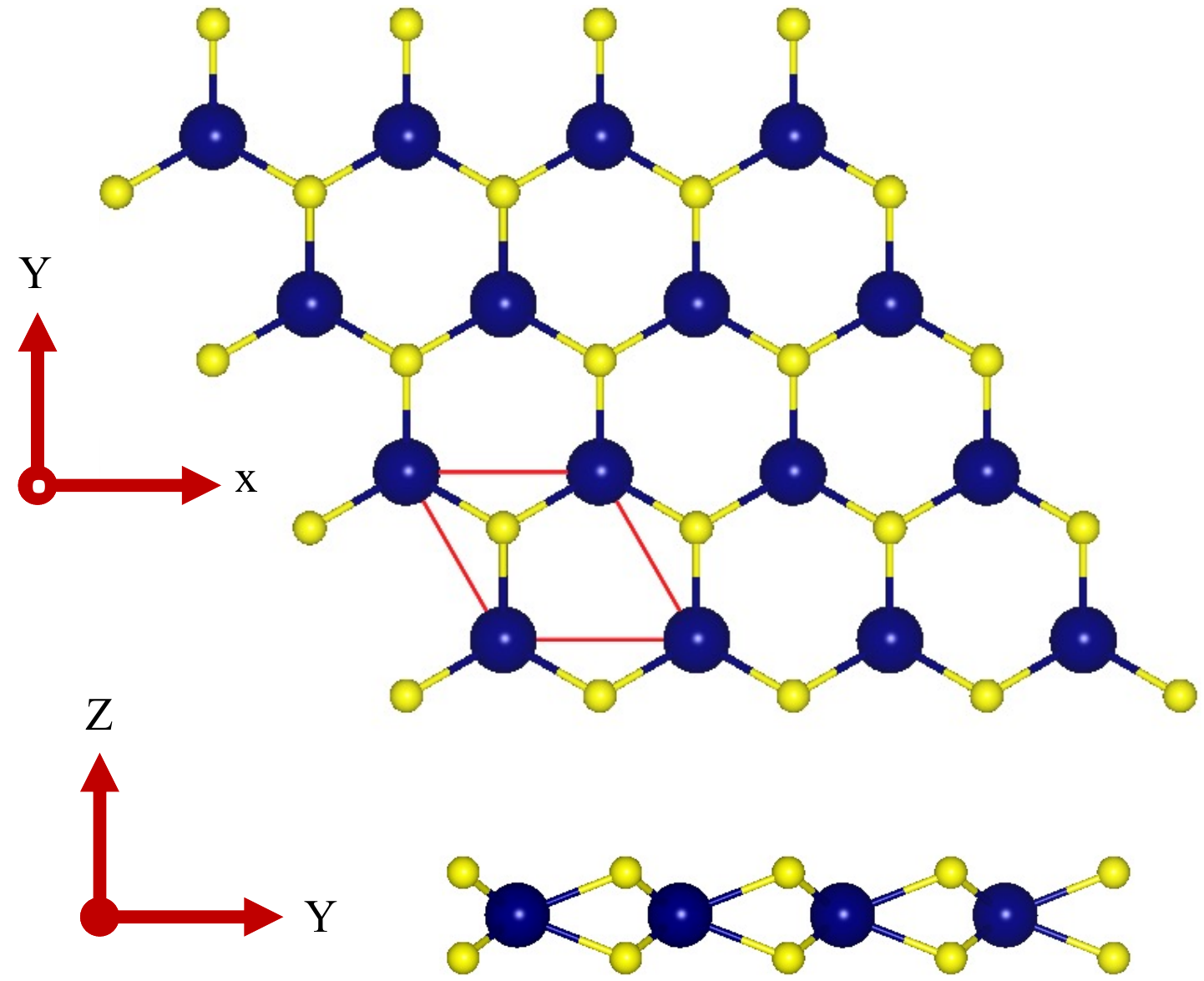}
    \caption{Top and side view of the H-phase monolayer EuS$_2$. The dark blue balls and yellow balls represent Eu and S atoms, respectively.  The definition of the axes is shown with arrows. 
    }
    \label{fig:atomic_structure}
\end{figure}

\begin{figure*}[ht]
    \includegraphics[angle=0, width=0.96\textwidth]{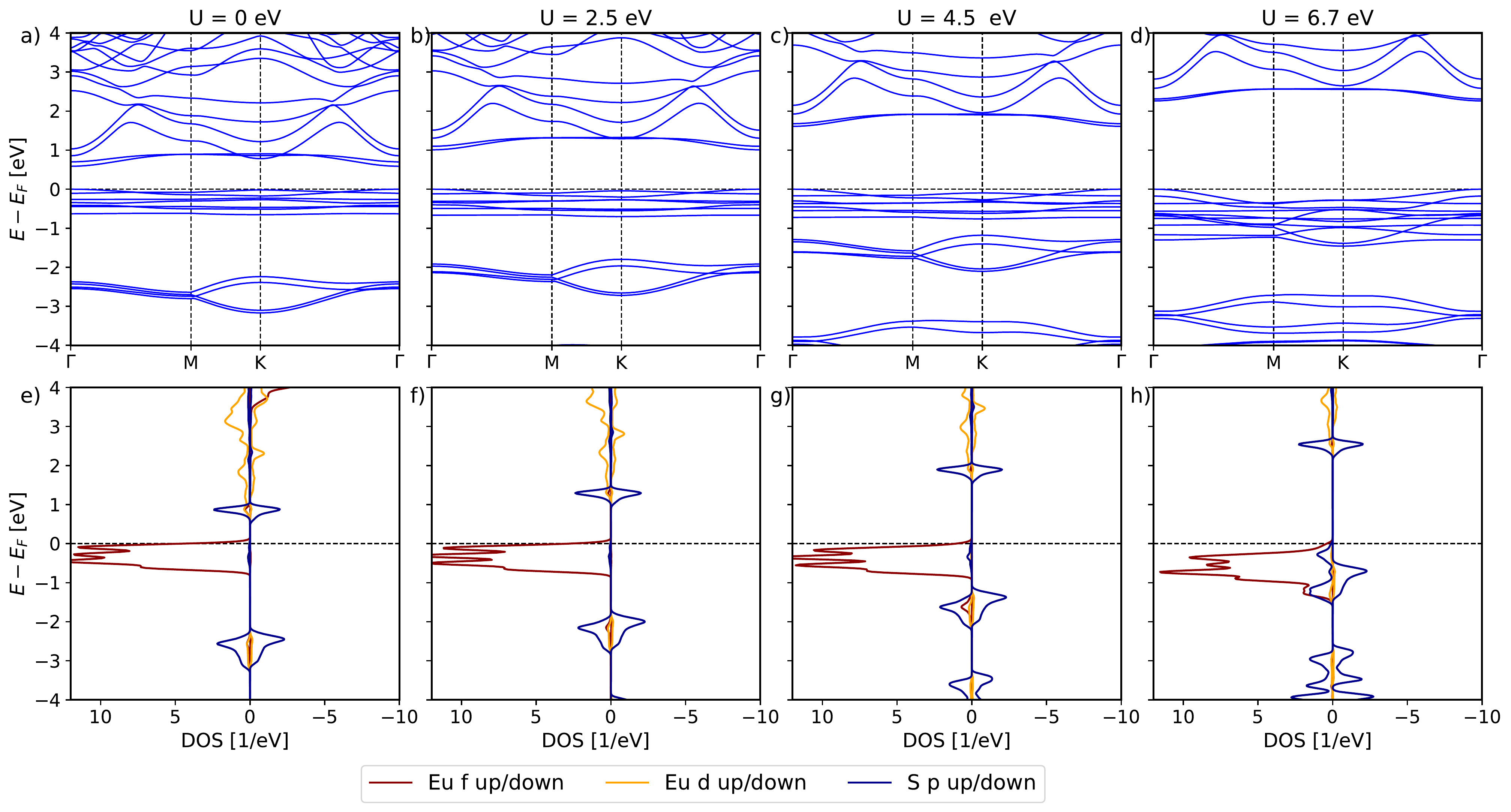}
    \caption{Electronic structure of H-phase  $\text{EuS}_2$ monolayer as a function of Coulomb repulsion strength $U$. In (a)-(d) the band structures are shown for $U=0$, $2.5$, $4.5$ and $6.7\ \mathrm{eV}$, respectively. In (e)-(h) the corresponding evolution of the spin-resolved density of states is shown.
    In (e)-(h) left and right parts of the figures correspond to the majority and minority spins, respectively, while 
    the density of states of Eu-$f$, Eu-$d$ and S-$p$ states is shown with dark red, orange and blue thin lines, respectively. While the dominance of the majority Eu-$f$ states at the Fermi level is evident, increasing $U$ drives the hybridization of Eu-$f$-states with S-$p$  states. This has drastic influence on the transport properties of the system, as discussed in the main text.}
    \label{fig:banddos}
\end{figure*}

\section {Results and discussion}
\subsection{Electronic structure of EuS$_2$}

We first discuss the electronic structure of EuS$_2$. We recall that consistent with Hund’s first rule the intratomic exchange interaction arranges the electronic configuration of the half-filled Eu $4f$ shell in the 4$f^7$ configuration, giving rise to a large magnetic moment of 6.95\,$\mu_\mathrm{B}$ $-$ a value which is robust with respect to the correlation strength. This is significantly larger than the spin moments in other TMD materials, compare e.g.~to Refs.~\cite{Ataca2012,Ma2012,Kan2014}. 
 In Fig.~\ref{fig:banddos} we show the  computed band structure of the system, (a-d), together with the densities of states (DOS), (e-h), for the values of $U$ of 0\,eV (a,e), 2.5\,eV (b,f), 4.5\,eV (c,g) and 6.7\,eV (d,h).
The quantitative features of the electronic structure of  EuS$_2$ are similar irrespective of $U$: The highest occupied states are formed by the Eu $4f$ majority electrons. The valence band maximum is positioned at the $\Gamma$ point. We find $p$ states of the chalcogen S atoms well separated and lying above and below the $f$ states. Together with Eu $f$ and $d$ states the $p$ states of S form bonding and antibonding states that have an energy separation of about 3.5 eV irrespective of the spin-channel. The lowest conduction band is relatively flat and it is formed by the states with primarily S-$p$ character.
%[\color{red}I think these might be p-sigma (surface) states in the atomic plane of S atoms, could be analysed ]
The conduction band minimum is also positioned at the $\Gamma$ point and EuS$_2$  monolayer exhibits a direct band gap of $f$-$p$ transition. The higher conduction bands following the S-$p$ band are formed by highly dispersing Eu 6$s$ and 5$d$ states. The localized occupied 4$f$ states act as a repulsive scattering potential to the S-$p$ electrons which
moves bonding and antibonding S-$p$ states in the majority and minority spin-channel to lower and higher energy, respectively.
This introduces a small exchange splitting, visible in Fig.~\ref{fig:banddos}(a)-(d), and a magnetic moment of $-0.28$\,$\mu_\mathrm{B}$ on S-atoms. The values of the orbital moments of Eu and S atoms are practically  negligible.

Focusing on the 4$f$ states, in the ground state, at $U = 0$\,eV, the majority $f$-states are separated from the conduction S $p$-states by about 1\,eV (direct band gap of 0.59\,eV at $\Gamma$), see Fig. 2(a,e). Increasing $U$ leads to an increase of the $f$-$p$ gap without a change of dispersion, and the 4$f$ states move down in energy closer to the occupied S $p$-valence states (or, alternatively, since the Fermi energy is fixed by the band edge of the majority $4f$ states, the S $p$-states move up in energy with respect to the $f$-states). At $U=6.7$\,eV this results in a $p$-$f$ energy separation of about 3\,eV at the Fermi energy (direct band gap of 2.27\,eV at $\Gamma$), Fig.~\ref{fig:banddos}(d,h). This puts EuS$_2$ among the large band gap TMD semiconductors~\cite{Guo2014,Jiao2016}. On the other hand, increasing $U$ drives the lower $p$-states into the region of majority $f$-states. In turn, this drives a strong hybridization between the $p$-, $d$- and $f$-states, which result in the orbital complexity and strong modifications that $U$ brings to transport properties of the doped system which we discuss below.
We conclude at this point that the EuS$_2$ monolayer is a ferromagnetic semiconductor with a direct band gap, whose value depends on the choice of $U$. The studied system  can be turned into a half-metal upon replacing Eu by e.g. Gd forming an Eu$_{1-x}$Gd$_x$S$_2$ layer.

\begin{figure*}[t!]
    \includegraphics[angle=0, width=0.58\textwidth]{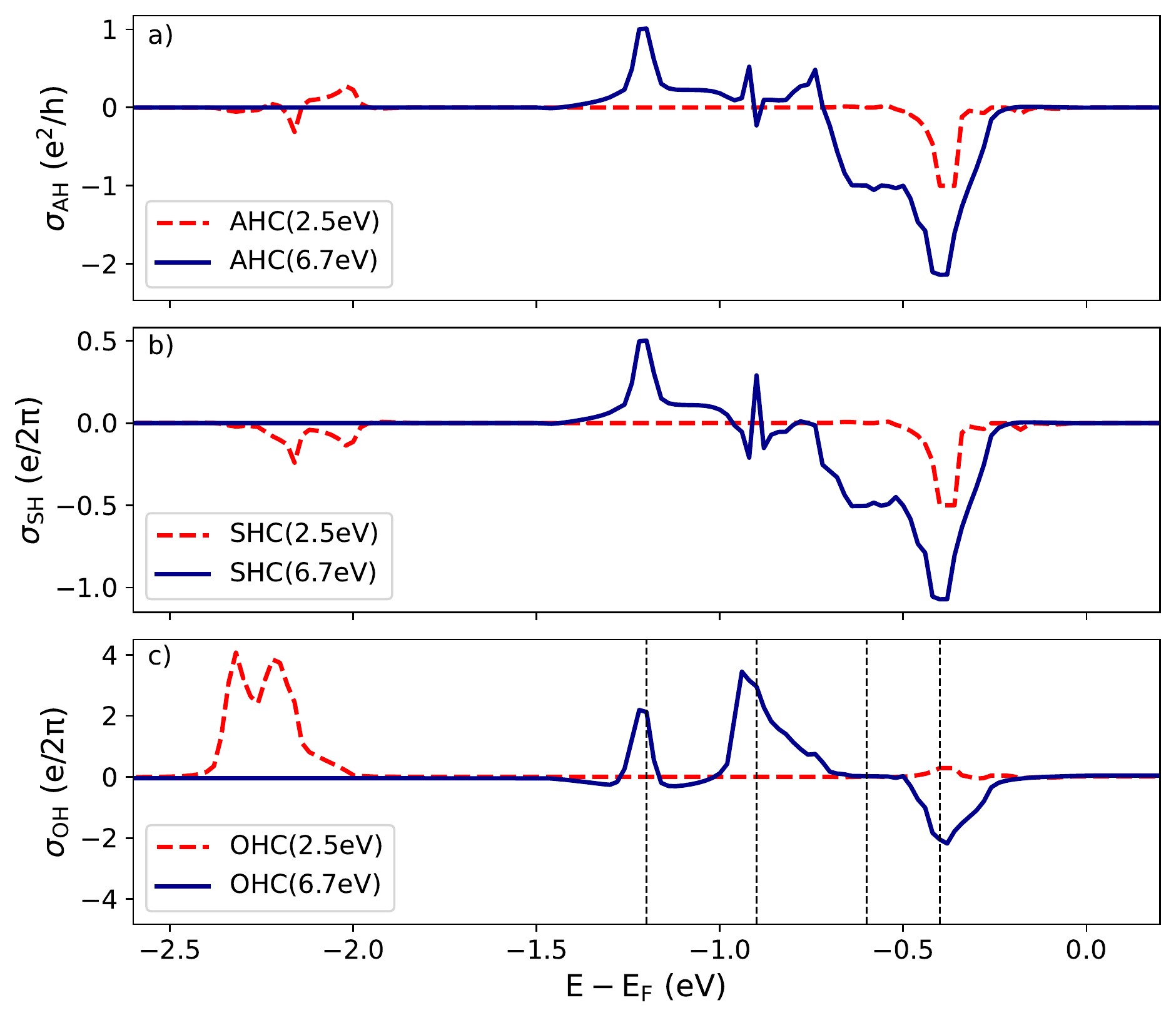}
    \caption{Transport properties of monolayer EuS$_2$ as  a function of band filling.
    (a) Anomalous Hall conductivity (AHC) $\sigma_{\rm AH}$, (b) spin Hall conductivity (SHC) $\sigma_{\rm SH}$, and (c) orbital Hall conductivity (OHC) $\sigma_{\rm OH}$, computed as a function of band filling in EuS$_2$ for $U=2.5$\,eV (red dashed line) and 6.7\,eV (blue solid line). Vertical dashed lines in (c) are guides to the eye marking the position of energy at $-1.2$, $-0.9$, $-0.6$ and $-0.4$\,eV. 
    }
    \label{fig:anomalous}
\end{figure*}

\subsection{Anomalous Hall effect in EuS$_2$}

Next, we proceed to investigate the AHE in the system.
%.  To study the transport propertied of H-EuS$_2$. We proceed to investigate AHC which is an intriguing transport phenomenon occurring typically in ferromagnets as a consequence of broken time reversal symmetry and spin-orbit interaction
Here,  we focus on the intrinsic Berry curvature contribution to the AHC tensor~\cite{Nagaosa2010}. 
%employing the $\boldsymbol {k}$-point band interpolation technique. 
To compute the Berry curvature from maximally-localized Wannier functions we first construct a tight-binding MLWFs Hamiltonian, and use the Wannier interpolation technique to efficiently evaluate the $xy$-component of the  AHC, $\sigma_{\rm AH}$,  as a Brillouin zone integral  on a 300$\times$300 mesh of $k$-points according to the expression: 
%projected from the GGA+U+SOC Bloch wavefunctions. Specifying all $p$ orbital state of  S atom and all $d$,$f$ orbital states of Eu atom as well. Defining 36 WFs in total. 
%We used  $300\times 300$ $\boldsymbol {k}$-mesh points in the 2D zone to compute the AHC. 
%The main equation for intrinsic AHC that expressed as a function of the integral of total Berry curvature $\Omega_\text{xy}$,
%
\begin{equation}
\sigma_{\rm{AH}}=-\frac{e^{2}}{\hbar}\sum_n\int_\text{BZ} \frac{d^2\boldsymbol{k}}{(2\pi)^2}f_{n\boldsymbol{k}}\,\Omega_{n\boldsymbol{k}},
\label{eq:anomloushall}
\end{equation}
with $f_{n\boldsymbol{k}}$ as the Fermi-Dirac distribution function, and the Berry curvature $\Omega_{n\mathbf{k}}$ of a state $n$ at point $\boldsymbol{k}$ as given by
\begin{equation}
\Omega_{n\boldsymbol{k}}=2\hbar^{2} \sum _{n\neq m}
\mathrm{Im}
\left[
\frac{\left<u_{n\boldsymbol{k}} \left|v_{x}\right|u_{m\boldsymbol{k}}\right>\left< u_{m\boldsymbol{k}}\left|v _{y}\right|u_{n\boldsymbol{k}}\right>}{(E_{n\boldsymbol{k}}-E_{m\boldsymbol{k}}+i\eta )^2}
\right],
 \end{equation}
where $E_{n\boldsymbol{k}}$ is the energy of a Bloch state with lattice periodic part of Bloch wave function given by $u_{n\boldsymbol{k}}$,  
and $v_{i}$ is the $i$'th Cartesian component of the velocity operator. For improving the convergence, we set $\eta=25\ \mathrm{meV}$.

%\begin{eqnarray}
%\ket{\psi_{n\mathbf{k}}}
%\\
%\bra{\psi_{n\mathbf{k}}}
%\\
%\braket{\psi_{n\mathbf{k}}| A | \psi_{m\mathbf{k}}}
%\end{eqnarray}

%\begin{figure}[ht!]
%    \includegraphics[angle=0, width=0.48\textwidth]{Fig3.pdf}
%    \caption{Anomalous Hall effect in $\text{H-phase EuS}_2$.
%    (a) Anomalous Hall conductivity of $\text{H-phase EuS}_2$, $\sigma^{z}_{\rm AH}$, as a function of Fermi energy calculated with SOC for $U$ of 2.5 and 6.7\,eV. (b-c) The band structure of $\text{H-phase EuS}_2$ just below the true Fermi level for $U$ of 2.5\,eV (b) and 6.7\,eV (c).
%    (d-e) Berry curvature  plotted along high symmetry lines in the Brillouin zone at different Fermi energy positions indicated with black and blue dashed lines in (b-c). For a discussion see main text.}
            
%    \label{fig:anomalous}
%\end{figure}

The results of our calculations of the AHC are summarized in Fig.~\ref{fig:anomalous}(a) for the cases of $U=2.5$ and 6.7\,eV, of which we first discuss the former. 
The energy dependence of the AHC for $U=2.5$\,eV is shown with a dashed line in Fig.~\ref{fig:anomalous}(a), where we observe that the AHC is non-zero only in a narrow window between $-0.6$ and $+0$\,eV, and $-3$ to $-2$ eV. This corresponds to the upper region of majority $f$-states, and the region of occupied S $p$-states, respectively. The out-of-plane spin and orbital polarization of these states is shown in Figs.~\ref{fig:Orbital}(a,b). We observe that the low lying $p$-states are slightly exchange split (by 50$-$100\,meV), perfectly spin-polarized, and they display very strong orbital polarization, especially at the top and bottom of the group. This is consistent with the process of $p_x\pm p_y$ orbital polarization,
confirmed by  the DOS analysis (not shown), giving rise to non-trivial Chern numbers and a non-vanishing AHC in the region of these states~\cite{Zhang2013}.
By inspecting the region of $f$-states in Fig.~\ref{fig:Orbital}(a,b) closer, we observe a formation of relatively large gaps among the groups of $f$-states, separated by the combined effect of crystal field and spin-orbit interaction. The topological  nature of these gaps is predominantly trivial,~i.e.~the quantized value of the AHC within these gaps is topologically trivial, except for the case of the gap at about $-0.4$\,eV, where the AHC is quantized to a value of $-1\,\frac{e^2}{h}$ (i.e.~having a Chern number of $-1$). From the analysis of the orbital character of contributing states it is clear that this is 
the region where the overall change in the sign of orbital polarization of the states around the gap takes place. This change in the orbital polarization across the gap mediated by the spin-conserving part of spin-orbit interaction can be closely associated with the non-trivial AHC in this region~\cite{Zhang2013}. 
%\begin{figure}[t!]
% \includegraphics[angle=0, width=0.48\textwidth]{Fig5.pdf}
%\caption{Out-of-pane spin (right) and orbital (left) %texture in EuS$_2$ calculated with SOC for different parameters: (a-b) for $U=2.5$\,eV and position of $E_F$ at $-0.4$\,eV; (c-d) for $U=6.7$\,eV and position of $E_F$ at $-0.5$\,eV; and (e-f) for $U=6.7$\,eV and position of $E_F$ at $-0.4$\,eV. The overall difference in sign of the textures in (a) and (b) can be observed. Note that the scale, indicated with a color bar, is different for all cases.}
%\label{fig:Orbital}
%\end{figure}

At this point we would like to comment on the logic behind energy-resolved orbital polarization of the system for $U=2.5$\,eV. For this correlation strength, the Eu-$f$ states are separated from the $p$-states of S atoms. The  $4f$ electrons are localized, and experience a non-spherical potential  exhibiting the point group symmetry of the Eu atom embedded into EuS$_2$ monolayer. Therefore, the development of the orbital moments is a result of a competition between Coulomb interaction, summarized under Hund’s second rule, and the crystal field. We notice that our results still resemble Hund’s second and third rule,~Fig.~\ref{fig:Orbital}(a,b). In the lower half of the $4f$-majority states, the orbital moment is opposite to the spin moment,  while for the upper half of the $4f$ states the orbital moment is parallel to the spin moment, which results in a neglibile orbital moment when summed over all majority $4f$ states. For S-$4p$ states the situation is very different since the $p$-states are subject to a large band dispersion where crystal field effects dominate. For states along $\Gamma$-M  the dispersion of the S $p$-states is rather small and Hund’s third rule is obeyed since the orbital moments of the lower (upper) half of the bands is antiparallel (parallel) to the spin moments, but the strong crystal field associated with the band dispersion along the high-symmetry lines M-K-$\Gamma$ makes Hund’s third rule inapplicable. This violation becomes even more prominent for larger values of the correlation strength.

The effect of direct interaction between the $p$-states of S and $f$-states of Eu, taking place when we increase the correlation strength to $U=6.7$\,eV has a profound effect on the AHE.
From the analysis of the DOS and spin-polarization of the states shown in Fig.~\ref{fig:Orbital}(c) we observe that minority $p$-states are easily identifiable as they do not at all hybridize with the $f$-states of opposite spin. The hybridization of the majority $p$- and $f$-states is, on the other hand, drastic, which is evidenced by strong changes in the position and dispersion of the bands, as well as increased spread of Eu and S states over the region of almost 1.5\,eV in energy for this spin channel. The resulting orbital complexity of hybridized states, visble in Fig.~\ref{fig:Orbital}(c) has an amplifying effect on the overall magnitude of the AHC. In particular,  close to quantized values of the AHC are achieved for the lower bands around $ -1.2$\,eV  ($\sigma_{\rm AH}\approx +1\,\frac{e^2}{h}$ related to a global gap present there, crossed by minority-S states), a wide plateau around $-0.6$\,eV (with $\sigma_{\rm AH}\approx -1\,\frac{e^2}{h}$, related to the non-trivial gap between three upper and five lower majority bands, separated by an ``inactive" band at $-0.55$\,eV), as well as a peak at $-0.4$\,eV (with $\sigma_{\rm AH}\approx -2\,\frac{e^2}{h}$).

\subsection{Spin Hall effect in EuS$_2$}

%{\bf CITE Kontani et al. PRL 2008.}
We now move on to the study of SHE and OHE in the system. While the SHE has been explicitly studied in group-VI dichalcogenides \cite{Feng2012,Xiao2012}, very recently, Canonico and co-workers investigated the orbital Hall effect in non-magnetic H-phase 2D TMDs MoS$_2$ and WS$_2$~\cite{Canonico2020,Canonico2020*}, addressing especially a question of quantization of the OHC within the electronic energy gaps, and  corresponding emergence of the  quantum orbital Hall insulating phase. %their findings presented these systems are orbital Hall insulators and both host a sizeable plateaus OHE within their electronic energy gaps. 
The study of the OHE is a subject of general interest since the magnitude of the OHC can be often  dominant over that of the SHC, which suggests systems with large OHE as potential platform for novel concepts in orbitronics~\cite{kontani2008,Go2021}.
%It was also speculated that large magnitude of orbital Hall effect compared to SHC, renders these materials interestingly favourable to study the OHC. 

\begin{figure}[t!]
 \includegraphics[angle=0, width=0.48\textwidth]{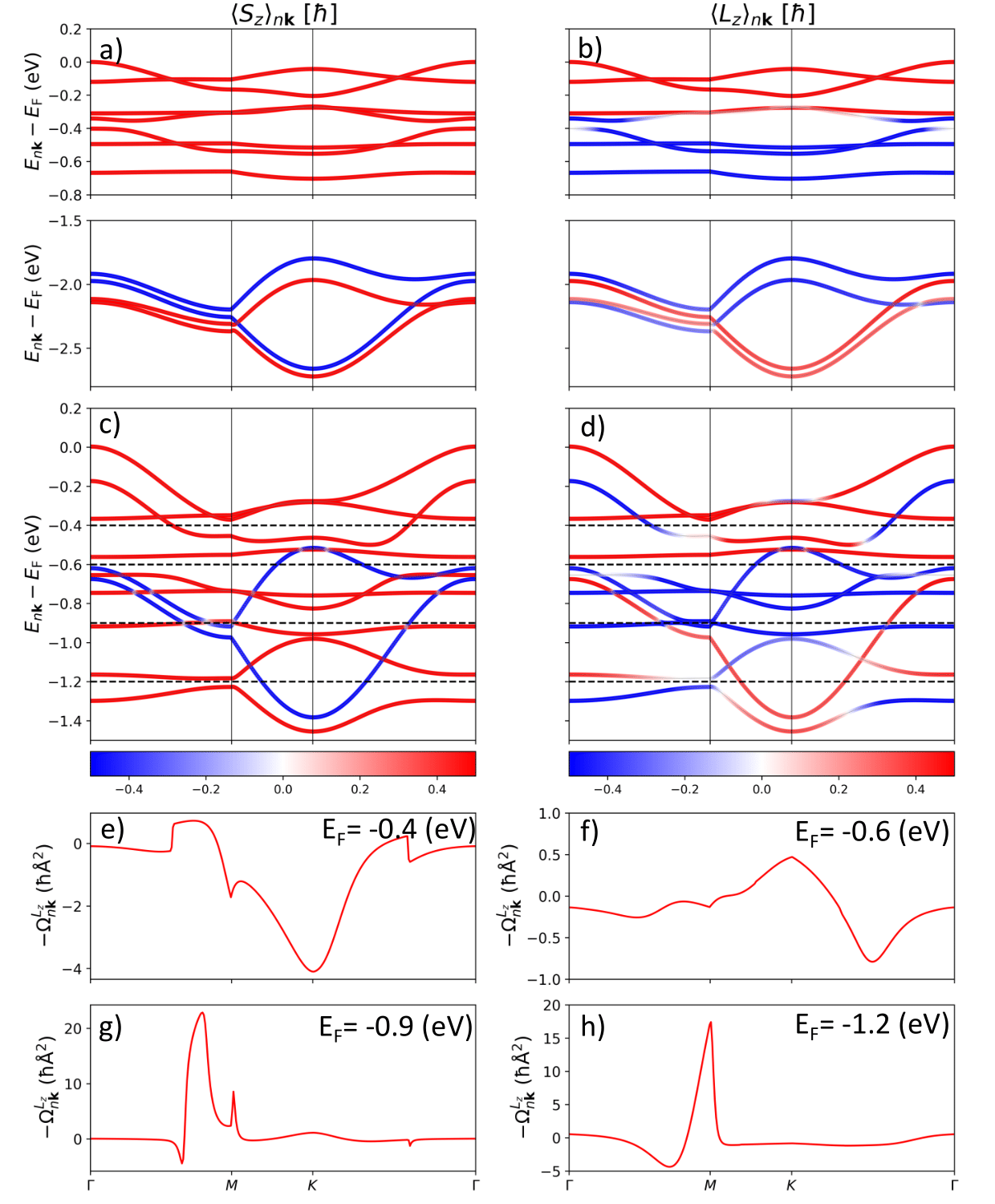}
\caption{Evolution the out-of-plane of spin, $\langle S_z\rangle_{n\mathbf{k}}$ (left panel), and orbital polarization, $\langle L_z\rangle_{n\mathbf{k}}$ (right panel), of occupied states in EuS$_2$. (a) The majority $f$-states of Eu (upper panel) and S $p$-states (lower panel) for the case of $U=2.5$\,eV (c), and of the hybridized group of $f$ and $p$-states for the case of $U=6.7$\,eV. (b,d): Analogous to (a,c). The value scales are indicated by the color bars at the bottom in units of $\hbar$. Horizontal dashed lines in (c-d) are guides to the eye marking the position of energy at $-1.2$, $-0.9$, $-0.6$ and $-0.4$\,eV. (e-f) Distribution of the orbital Berry curvature along high symmetry lines for the indicated position of the Fermi energy.
} \label{fig:Orbital}
\end{figure}

Motivated by these considerations, we proceed to explore the possibility that EuS$_2$ can host SHE and OHE. Following the established methodology~\cite{Go2018,Go2021*}, we calculate the SHC and OHC, which mediates a generation of a transverse in-plane current of out-of-plane (spin or orbital) angular momentum in response to an applied electric field, according to the Kubo expression:
\begin{equation}
\sigma_{\text{OH/SH}} =
\frac{e}{\hbar}
\sum_{n}\int \frac{d^{2}\boldsymbol{k}}{(2\pi)^{{2}}}f_{n_{\boldsymbol{k}}}\,\Omega_{n\boldsymbol{k}}^{J_z},
\end{equation}
where the so-called spin (orbital) Berry curvature reads as:
\begin{equation}
\Omega_{n\boldsymbol{k}}^{J_z}=
2\hbar^{2}\sum_{m\neq n} 
{\rm Im}
\left[
\frac{
\left \langle 
u_{n\boldsymbol{k}}
| j_{y}^{J_z}| 
u_{m\boldsymbol{k}}
\right\rangle
\left \langle  
u_{m\boldsymbol{k}} 
|v_x| 
u_{m\boldsymbol{k}} 
\right \rangle
}
{
(E_{n\boldsymbol{k}} - E_{m\boldsymbol{k}}+i\eta )^2
}
\right]
,
\end{equation}
with $j_{y}^{J_z}$ as the spin ($J_z=S_z$, the $z$ component of the spin operator) or orbital ($J_z=L_z$, the $z$ component of the local angular momentum operator) current operator defined as $j_y^{J_z} = (v_y J_z + J_z v_y)/2$.

We first look at the SHE, presenting the results of the calculations in Fig.~\ref{fig:anomalous}(b). For the case of $U=2.5$\,eV we find that for the $f$-states the SHC precisely follows the AHC as the band filling in varied. This is consistent with the picture of the spin angular momentum carried by the anomalous Hall current of fully spin-polarized $f$-bands in that energy region. The behavior of the SHC of S $p$-states at lower energies is more complicated, however, the consistently negative sign of the SHC can be understood from the sign reversal  both in net spin-polarization as well as in the AHC at the top and at the bottom of the S-states. Remarkably, an almost perfect one-to-one correspondence between the energy dependence of the SHC and AHC still persists  for a more complex case of $U=6.7$\,eV. This can be again understood taking into account pure spin character of $f$-states. The presence in that energy region of minority S-states, see Fig.~\ref{fig:Orbital}(c-d), is the reason for a discrepancy in the shape of the SHC and AHC between $-1.0$ and $-0.6$\,eV, which can be explained by taking into account a small negative spin Hall signal that the minority S-states bring with them into that energy region.

\subsection{Orbital Hall effect in EuS$_2$}

Finally, we turn to the analysis of the OHE in the system. We present the results of our calculations of the OHC in Fig.~\ref{fig:anomalous}(c). We first observe that in the
``unhybridized" case of $U=2.5$\,eV, the OHC of $f$-states is minimal, reaching in magnitude the SHC in the region around $-0.4$\,eV, where the AHE and SHE are present as well. Most remarkable is the gigantic OHE carried by the $p$-states of S atoms, positioned between $-3$ and $-2$\,eV, Fig.~\ref{fig:Orbital}(b). The OHC in this region is almost one order of magnitude larger than the SHC or OHC of the $f$-states for this value of $U$. By inspecting the correlation between the electronic structure, orbital polarization and orbital Hall conductivity, we can identify a region of energy between $-2.4$ and $-2.1$\,eV as the dominant source of the OHE due to contributions coming from crossings among bands of different orbital character, clearly visible close to the M-point and the $\Gamma$-point in Fig.~\ref{fig:Orbital}(b), as well as larger parts of the Brillouin zone where there is strong $k$-dependent exchange of orbital angular momentum among the oppositely-polarized bands, which is a direct consequence of $k$-dependent hybridization strength.  

Upon increasing the value of $U$ to 6.7\,eV the complicated $p-d-f$ hybridization taking place among S and Eu atoms exerts a striking influence on the properties of the OHE of the $p-f$ group of states below the Fermi energy. This is reflected in a three to four times larger magnitude of the OHC as compared to the SHC observed in the region of majority $f$-states, Fig.~\ref{fig:anomalous}(c), and, in contrast to expectations from oversimplifying band-filling arguments applied in the past to non-magnetic materials~\cite{Tanaka2008},  a very non-trivial dependence of the OHC on the Fermi energy. At first, it is tempting to attribute the large computed OHC to S $p$-states, which promote a strong OHE already for smaller values of $U$ without any need for aid from the side of $f$-orbitals. However, a closer analysis shows that this is not the case. First, we take a look at the region of energy around $-1.2$\,eV, where the lowest of the peaks in the OHC appears [this energy value is indicated with dashed lines in Fig.~\ref{fig:anomalous}(c) and Fig.~\ref{fig:Orbital}(c-d)]. Here, the large peak in the OHC originates from an anticrossing among two strongly Eu-S hybridized bands [see also the corresponding DOS in Fig.~\ref{fig:banddos}(h)], positioned at that energy, where an exchange of the orbital polarization is particularly visible around the M-point.
A very similar situation is encountered next to and directly at the M-point around $E_F=-0.9$\,eV, where the crucial role of the orbital angular momentum exchange and large values of the OHC displayed by the majority $p$-states become very apparent. The distribution of the orbital Berry curvature of occupied states up to the corresponding Fermi energy, shown in Fig.~\ref{fig:Orbital}(g-h), confirms this picture.

Despite a sizable magnitude of the AHC and SHC in the region of energy between $-0.7$ and $-0.5$\,eV [the value of $-0.6$\,eV is marked with dashed lines in Fig.~\ref{fig:anomalous}(c) and Fig.~\ref{fig:Orbital}(c-d)], the OHC is suppressed there, which is quite unexpected. The reason for this is the absence of the $k$-dependent exchange of the orbital polarization among pairs of bands, which appear in this energy window. And although there is a crossing of positively and negatively orbitally polarized bands around $-0.55$\,eV in the vicinity of the K-point, no hybridization or exchange of orbital angular momentum takes place there. The corresponding orbital Berry curvature for this Fermi energy, shown in Fig.~\ref{fig:Orbital}(f), is overall more than one order of magnitude smaller than for the previously considered cases. This solidifies the $k$-dependent orbital polarization exchange as a driving force behind large OHE observed in the studied system at lower energies. 

The scenario for the large OHC in the vicinity of $-0.4$\,eV is different: here, it is the orbital exchange around M and especially K, taking place among pairs of bands well-separated in energy (i.e. the 2nd and 3rd bands when counting down from zero energy) which gives rise to moderate values of the Berry curvature exhibited over larger regions of the reciprocal spaces, which is responsible for the overall sizeable OHC.
The negative sign of the OHC here, which cannot be achieved neither with $p$ nor $f$-states separately (for lower $U$),
emphasizes once more the importance of the $p$-$f$ hybridization for the orbital Hall physics in EuS$_2$ in particular, and in rare earth dichalcogenides in general.

\section{Conclusions}
In this work, we made first steps in exploring the prospects of rare earth dichalcogenides as a source of transverse topological currents of charge, spin and orbital angular momenta. We chose EuS$_2$ as a representative of this class of materials, and computed its structural, electronic, magnetic and transport properties from first principles. While expectedly finding that EuS$_2$ exhibits pronounced magnetism, we also found that this $f$-material gives rise to very strong Hall responses and hosts topologically non-trivial bands, despite strong correlation among  electrons in the $f$-shell. We identified that the reason for this is a strong degree of $p-d-f$ hybridization among the Eu and S atoms, enhanced by stronger correlation effects owing to peculiar electronic structure of EuS$_2$. Effectively, S-originated $p$-states and Eu $f$-states unite their efforts in enhancing each other's anomalous and spin Hall response. As a result of complex orbital hybridization among the latter states, the composite $p-f$ group of states also exhibits a very large orbital Hall effect, which is dominant over the SHE by about a factor of three. We identify $k$-dependent orbital exchange among pairs of hybridized bands as the major source of the large orbital Hall effect. Our results not only promote rare-earth dichalcogenides as an exciting platform for spin and orbital physics, but also advance our understanding of complex orbital transport manifestations of correlated magnetic two-dimensional materials.       

%Using the first principles calculation, we have calculated the electronic structure, in addition to the anomalous, spin and orbital Hall conductivity of H-phase EuS$_2$ monolayer. We find that H-EuS$_2$ is a ferromagnetic semicoductor with high magnetic moment of around $6.9\ \mu_B$ per the unit cell and direct band gap of $2.27\ \mathrm{eV}$ when $6.7\ \mathrm{eV}$ $U_\text{eff}$ is applied. The strong effect of applying $U_\text{eff}$ is clearly shown. On another hand, the present calculations indicate that theoretical AHC is large as a function of Fermi energy positions.
%Finally our result demonstrates that the OHC is an order of magnitude larger compared to the SHC. That characterize this  material as interesting platform for the orbitronics application.
%\cite{Go2021,kontani2008}.

\begin{acknowledgements}
This work was supported by the Federal Ministry of Education and Research of Germany in the framework of the Palestinian-German Science Bridge (BMBF grant number 01DH16027). We also gratefully acknowledge financial support by the Deutsche Forschungsgemeinschaft (DFG, German Research Foundation) $-$ TRR 173/2 $-$ 268565370 (projects A11 and A01), CRC 1238 - 277146847 (Project C01), and the Sino-German research project DISTOMAT (MO 1731/10-1).  We  also gratefully acknowledge the J\"ulich Supercomputing Centre and RWTH Aachen University for providing computational resources under projects jiff40 and jara0062. 
%{\bf \color{red} jiff40?} 

\end{acknowledgements}

\bibliography{bib_EuS2}

\end{document}